\documentclass[reprint, superscriptaddress, aps, prl]{revtex4-1}

\usepackage[T1]{fontenc} 

\usepackage{url}
\usepackage[
  unicode,
  pdfusetitle,    
  pdfcreator={},
  pdfproducer={}, 
]{hyperref}

\usepackage{mathtools}
\let\theta\vartheta
\newcommand*{\dd}{\mathrm{d}}
\newcommand*{\pp}{\mathrm{p}}
\newcommand*{\bu}{\textbf{u}}
\newcommand*{\br}{\textbf{r}}
\newcommand*{\bq}{\textbf{q}}

\usepackage{xcolor}

\colorlet{myred}{red!80!black}

\begin{document}

\title{Connectivity, Not Density, Dictates Percolation in Nematic Liquid Crystals of Slender Nanoparticles}
\author{Shari P. Finner}
\email{s.p.finner@tue.nl}
\affiliation{Department of Applied Physics, Eindhoven University of Technology, P.O. Box 513,
3500 MB Eindhoven, The Netherlands}
\author{Tanja Schilling}
\email{tanja.schilling@physik.uni-freiburg.de}
\affiliation{Physikalisches Institut, Albert-Ludwigs-Universit{\"a}t Freiburg, 79104 Freiburg, Germany}
\author{Paul van der Schoot}
\affiliation{Department of Applied Physics, Eindhoven University of Technology, P.O. Box 513,
3500 MB Eindhoven, The Netherlands}
\affiliation{Institute for Theoretical Physics, Utrecht University, Princetonplein 5, 3584 CC Utrecht, The Netherlands}

\begin{abstract}	
	We show by means of continuum theory and simulations that geometric percolation in uniaxial nematics of hard slender particles is fundamentally different from that in isotropic dispersions.
	In the nematic, percolation depends only very weakly on the density and is, in essence, determined by a distance criterion that defines connectivity.
	This unexpected finding has its roots in the non-trivial coupling between the density and the degree of orientational order that dictate the mean number of particle contacts.
	Clusters in the nematic are much longer than wide, suggesting the use of nematics for nanocomposites with strongly anisotropic transport properties.
\end{abstract}  

\maketitle

The mechanical and transport properties of polymeric materials can be enhanced significantly by the addition of nanoparticles.\,\cite{KoningCNTbook}
For nanoparticles to strongly affect material properties, a minimum filler fraction is required as they need to be, in some sense, connected and form a system-spanning network.
The electrical (and thermal) conductivity of engineering plastics, for instance, can be increased drastically by the addition of small amounts of carbon nanotubes during the fluid production stages.\,\cite{KoningCNTbook,Moniruzzaman2006}
The nanotubes in the composite need not actually touch in order for electrons to pass through the network.
It is sufficient for them to be within a certain (tunneling) range for effective charge transport to take place.\,\cite{Shklovskii2006,Ambrosetti2010}
Hence, electrical percolation in this kind of system is often treated as geometric percolation, where connections are defined by a cut-off distance called the connectivity criterion, with consistent predictions.\,\cite{Ambrosetti2010, Otten2009}
In view of potential applications in nanotechnology, in particular in optoelectronics, photovoltaics and electromagnetic interference shielding, it is not surprising that geometric percolation in particulate dispersions is attracting a lot of interest.\,\cite{Moniruzzaman2006}

Elongated particles are particularly suited for such applications because of their low percolation threshold, which arguably is due to their large contact volume, \textit{i.e.}, the volume one particle can trace out while remaining connected to a second particle.\,\cite{Balberg1984,Coniglio1977,Bug1985}
Such particles not only have a large contact volume, but also a large excluded volume, which drives a transition from the isotropic to the uniaxial nematic liquid-crystalline state at filler fractions comparable to the expected percolation threshold.\,\cite{Onsager1949}
If the connectivity range is sufficiently small, percolation may be pre-empted by the isotropic-nematic transition.\,\cite{Otten2012}

Whilst many facets of geometric percolation in isotropic dispersions have been investigated, including the roles of polydispersity \cite{Otten2009,Nigro2013,Chatterjee2014,Finner2018}, external fields \cite{Otten2012,Finner2018,Kale2016,White2009}, interactions \cite{Schilling2007,Kyrylyuk2008}, flexibility \cite{Kyrylyuk2008,Kwon2016}, tortuosity and particle shape \cite{Drwenski2017,Kwon2016}, very little seems to be known about percolation in liquid-crystalline phases of rod-like particles.
What is known, is that forced alignment increases the percolation threshold.\,\cite{Otten2012,Chatterjee2014,Finner2018,Kale2016,White2009}
Here, we show by means of connectedness percolation theory and Monte Carlo (MC) simulations that the coupling between the density of hard slender particles and their degree of spontaneous alignment makes percolation in the uniaxial nematic phase highly unusual and fundamentally different from that in isotropic dispersions.

The reason is that the mean contact volume of the particles depends on the degree of order, which itself is density-controlled.
As a result, percolation can be completely suppressed if the connectivity criterion is too small, irrespective of the filler fraction and particle aspect ratio (in the slender-rod limit).
If the connectivity exceeds a critical value, percolation occurs at \textit{all} densities in the nematic.
In a narrow region between these two regimes, we find percolation in the low-density nematic that, remarkably, is lost upon particle addition.
The reason is that particles align more strongly, which leads to larger surface-to-surface distances and fewer contacts.
According to our theory and simulations, sub-critical clusters are much longer along the director than perpendicular to it, reflecting the underlying symmetry of the nematic phase.
Cluster dimensions in both directions diverge at the same density, with the same critical exponent but different prefactors.
This suggests that in thin films or fibers, nematics may be used to create materials with anisotropic conductivity.\,\cite{Zamora2011}

We first briefly describe the ingredients of our model, which hinges on a combination of Onsager theory, describing how the orientational distribution  of particles responds to density variations, and connectedness Ornstein-Zernike theory describing the cluster structure.\,\cite{Coniglio1977}
Our particles are cylindrical, interact via harshly repulsive steric interactions, and are sufficiently slender for the second virial approximation to hold.
In this limit, the distribution function, $\psi(\bu)$, of the particle orientation, $\bu$, obeys the Onsager equation
$\ln\psi(\bu)=k-8\pi^{-1}c\langle|\bu\times\bu'|\rangle'$.
Here, the constant $k$ enforces normalization of the distribution, and $\langle\dots\rangle\equiv\int\dd\bu(\dots)\psi(\bu)$, with a similar definition $\langle\dots\rangle'$ for the primed variable.\,\cite{Onsager1949} 
The dimensionless particle concentration $c\equiv\pi L^2D\rho/4$ is defined in terms of the number density $\rho$, the particle diameter $D$ and the particle length $L\gg D$.

We solve the Onsager equation numerically by recursive iteration of the function $\psi(\bu)$ as described in Ref.\,\cite{vanRoij2005}, and obtain stable and metastable solutions (in the biphasic region).
At coexistence, implying equal pressures and chemical potentials, the concentration in the isotropic phase is 3.29, and in the nematic 4.19.\,\cite{Odijk1986review}
For comparison, we also obtain analytical predictions using Odijk's Gaussian approximation to the orientational distribution,
$\psi(\bu)=\psi(\theta)=c^2\exp(-2c^2\theta^2/\pi)/\pi^2$ for $0\leq\theta\leq\pi/2$ 
and $\psi(\bu)=\psi(\pi-\theta)$ for $\pi/2\leq\theta\leq\pi$.\,\cite{Odijk1986review}
Here, $\theta$ denotes the angle with respect to the nematic director,
and $\langle\theta^2\rangle\sim\pi/2c^2$ in equilibrium.
The Gaussian distribution is most accurate deep in the nematic phase, but also reasonably accurate near melting conditions. \cite{Odijk1986review}

The second ingredient of our theory is the connectedness Ornstein-Zernike (cOZ) equation for the pair connectedness function $P(\br, \bu,\bu')$, describing the probability of finding two particles with orientations $\bu$ and $\bu'$ at relative position $\br$ in the same cluster.
In Fourier space, it reads
$\hat{P}(\bq,\bu,\bu')=\hat{C}^+(\bq,\bu,\bu')+\rho\langle\hat{C}^+(\bq,\bu,\bu'')\hat{P}(\bq,\bu'',\bu')\rangle''$,
with $\hat{C}^+(\bq,\bu,\bu')$ the direct connectedness function and $\bq$ the wave vector.\,\cite{Coniglio1977}
The first term describes two test particles that are part of the same cluster and cannot be separated by the removal of a single, third particle.
The second term represents all other types of cluster containing at least one particle which, upon removal, separates the test particles.
The density enters the cOZ equation directly, but also indirectly through the orientational averaging.
We consider two particles \textit{directly} connected if their shortest surface-to-surface distance is smaller than a cut-off distance $\lambda$, so that,
within the second virial approximation
$\hat{C}^+=2L^2\lambda|\bu\times\bu'|j_0\left(\frac{L}{2}\bq\cdot\bu\right)j_0\left(\frac{L}{2}\bq\cdot \bu'\right)$
with $j_0(x)\equiv\sin x/x$, provided $|\bq|D\ll1$.\,\cite{Otten2012,Coniglio1977,Onsager1949}
In nanocomposites, $\lambda$ is the average tunneling distance of electrons, which depends on the dielectric constant of the polymeric host medium, while in aqueous dispersions, where ions are the charge carriers, it must be the Debye length.
\footnote{The percolation threshold of charge-stabilized nanotubes can be measured by dielectric spectroscopy.\,\cite{Vigolo2005}
The dielectric constant peaks at the percolation threshold on account of a diverging mean cluster size.}

The size and structure of clusters are described by the ``cluster structure factor''
$S(\bq)\equiv1+\rho\langle\langle\hat{P}(\bq,\bu,\bu')\rangle\rangle'$.\,\cite{Coniglio1977}
The weight-average cluster size is $S=S(\textbf{0})$, and cluster dimensions along and perpendicular to the director, 
$\xi_{\|}$ and $\xi_{\bot}$, can be obtained from the Lorentzian
$S(q_{\|},q_{\bot})/S=(1+\xi_{\|}^2q_{\|}^2+\xi_{\bot}^2q_{\bot}^2)^{-1}$,
valid for long wavelengths, with $q_{\|}$ and $q_{\bot}$ the wave numbers along and perpendicular to the director.
We pinpoint the percolation threshold by finding the conditions for which $S\rightarrow\infty$ and $\xi_{\|,\bot}\rightarrow\infty$.
This we do numerically by recursive iteration of the cOZ equation for the function $\langle\hat{P}(\bq,\bu,\bu')\rangle'$, and analytically by invoking a variational principle.\,\cite{vdSchoot1990}
For the latter we define a function 
$m(\bq,\bu)=\sqrt{\psi(\bu)}\big[1+\rho\langle\hat{P}(\bq,\bu,\bu')\rangle'\big]$
and construct the functional 
$F[m]=\int\!\dd\bu\big[\frac{1}{2}m^2(\bq,\bu)-\sqrt{\psi(\bu)}m(\bq,\bu)\big]
 -\frac{1}{2}\rho\int\!\dd\bu \int\!\dd\bu'm(\bq,\bu)K(\bq,\bu,\bu')m(\bq,\bu')$,
 which, if extremized, produces the cOZ equation for $\langle\hat{P}(\bq,\bu,\bu')\rangle'$.
 Here,
 $K(\bq,\bu,\bu')\equiv\sqrt{\psi(\bu)}\hat{C}^+(\bq,\bu,\bu')\sqrt{\psi(\bu')}$.
A sensible trial function is $m(\bq,\bu)=M\sqrt{\psi(\bu)}$, with $M$ a variational parameter.
Setting $\partial F/\partial M=0$, we obtain the mean cluster size 
$S^{-1}=M^{-1}=1-2L^2\lambda\rho\langle\langle|\bu\times\bu'|\rangle\rangle'$,
an exact result for randomly oriented and perfectly parallel particles.\,\cite{Drwenski2017,Kyrylyuk2008}

\begin{figure}
	\includegraphics[width=\linewidth]{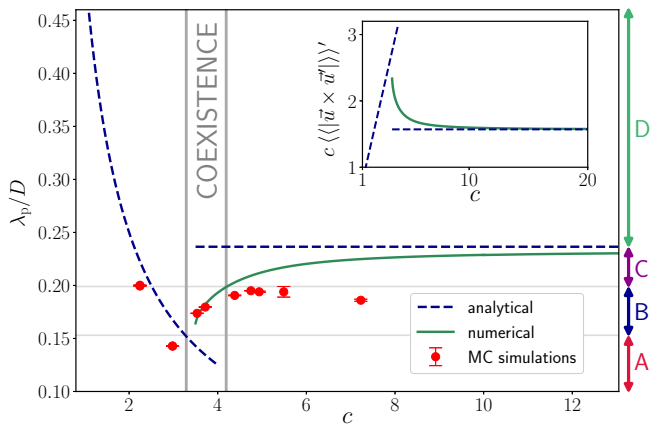}
	\caption{
		Critical connectivity $\lambda_\pp$ at percolation \textit{vs.} particle concentration $c$.
		Dashed lines present our analytical predictions in the isotropic phase (exact) and in the nematic (approximate).
		The solid line is our (exact) numerical solution.
		Filled circles indicate MC simulation results for $L/D=100$.
		Also indicated are four connectivity regimes according to our exact theoretical predictions.
		In regime A, the connectivity is too small for percolation to occur in any of the phases.
		In regime B, percolation is possible in the isotropic phase, but not in the nematic.
		In the regimes C and D, percolation occurs at sufficiently large concentrations in the isotropic phase.
		In the nematic, we have percolation at sufficiently \textit{low} concentrations in C, while in D percolation occurs irrespective of the concentration.
		\textbf{Inset:} Measure for the mean number of contacts per particle as a function of the concentration, explaining the four connectivity regimes (see the main text).
	}
	\label{fig:phasediagram}
\end{figure}

In the isotropic phase, $\langle\langle|\bu\times\bu'|\rangle\rangle'=\pi /4$, so $S^{-1}=1-2c\lambda/D$, producing a percolation threshold $c_\pp=D/2\lambda$, which agrees quantitatively with computer simulations for $L/D\gtrsim300$,\,\cite{Schilling2015} a condition easily met by carbon nanotubes.\,\cite{Pasquali2017}
Even for aspect ratio 100, the agreement with MC simulations is almost quantitative, see Figure \ref{fig:phasediagram}.
In the nematic phase, $\langle\langle|\bu\times\bu'|\rangle\rangle'\sim\pi/2c$ within the Gaussian approximation to leading order in $c\gg1$,\,\cite{Odijk1986review} with the surprising result
$S^{-1}=1-4\lambda/D$.
This defines a critical connectivity $\lambda_\pp=D/4$ below which percolation does not occur, and above which there is percolation irrespective of the particle concentration.
Apparently, in the nematic, a concentration increase is compensated by a concomitant decrease in contact volume.
This is confirmed using the more sophisticated trial function $m(\bq,\bu)=\sqrt{\psi(\bu)}\big(M+\theta^2N\big)$. 
Setting $\partial F/\partial M=\partial F/\partial N=0$, we find $\lambda_\pp=0.2368\,D$.

The Gaussian approximation becomes less accurate near melting conditions, so the cancellation is not exact, as Figure \ref{fig:phasediagram} shows.
Our numerical results for the critical connectivity increase with the density, but do saturate to a constant $\lambda_\pp\approx 0.2365\pm0.0005\,D$, close to our analytical prediction.
Comparison with MC simulations for $L/D=100$ confirms the increase of the critical connectivity in the low-density nematic.
The numerical prediction for the metastable nematic (in the biphasic regime) agrees well with our simulation results. 
However, in our simulations, the nematic seems stable, not metastable. 
Also, we see a slight downturn of $\lambda_\pp$ for the largest simulated density.
In a follow-up article we show, using a Lee-Parsons-type renormalization of $\hat{C}^+$,\,\cite{Schilling2015} that these discrepancies arise from the finite aspect ratio of the particles in our simulations.

Looking at whether or not percolation can occur in the stable isotropic and the stable nematic phase, we define in Figure \ref{fig:phasediagram} four connectivity regimes according to our (exact) prediction for the percolation threshold.
For a narrow connectivity range, percolation in the isotropic phase does not lead to percolation in the nematic (regime B).
Below this range (regime A), percolation cannot occur in any of the phases.
Above this range, percolation is possible in both phases (regimes C and D).
In regime C, percolation in the isotropic phase occurs at sufficiently large concentrations, and in the nematic at sufficiently \textit{low} concentrations, \textit{i.e.}, it may be lost again deeper into the nematic.
In regime D, we always find percolation in the nematic, in agreement with our variational theory.

The question arises what lies at the root of the complex relation between density, connectivity and the occurrence of system-spanning clusters.
To attempt answering this question, we plot in the inset of Figure \ref{fig:phasediagram} the concentration dependence of $c\langle\langle|\bu\times\bu'|\rangle\rangle'$, which for a given connectivity is an intuitive measure for the mean number of contacts of a rod.\,\cite{Nigro2013}
The Figure shows that, close to the isotropic-nematic transition, the number of particle contacts in the isotropic phase must always be larger than that in the nematic, explaining regime B in the percolation diagram.
In the low-density nematic, the number of contacts \textit{decreases} with concentration due to increasing alignment, before saturating to a constant value.
This is contrary to what happens in the isotropic phase and explains regimes C and D in Figure \ref{fig:phasediagram}.
Regime A arises, because for values of the connectivity smaller than $0.152\,D$, the nematic becomes the stable phase before the concentration in the isotropic phase can grow large enough to induce percolation.

That there is a saturation value of the number of contacts means that our numerical solution for $\langle\langle|\bu\times\bu'|\rangle\rangle'$ scales as $1/c$ for large $c$.
This agrees with our result in the Gaussian approximation, indicated by the horizontal dashed line.
The reason is that, in the nematic, the numerical solution to the Onsager equation obeys the high-density scaling function $\psi(c\theta)$, which also holds for the Gaussian distribution.\,\cite{vanRoij1996}
In the low-density nematic, the contact volume is larger than predicted by the Gaussian distribution because particles are less strongly aligned.
This explains why deeper into the nematic we need a larger connectivity for percolation to occur.
Because of the Gaussian approximation, our analytical prediction misses regime C but does predict the regimes A, B and D.

Before discussing cluster dimensions, we briefly turn to our MC simulations.
The data presented in Figure \ref{fig:phasediagram} were produced in a cuboid simulation box of dimensions $220\,D\times220\,D\times880\,D$ for the nematic and $(220\,D)^3$ for the isotropic phase, with periodic boundaries.
To evaluate finite-size effects, we investigated box dimensions of $L_{x,y}=[220\,D,\ldots, 660\,D]$ and $L_z=[220\,D, \ldots, 1760\,D]$ in the nematic, with particle numbers from $3042$ to $45$,$378$.
In the isotropic phase, $L_{x,y,z}=[220\,D, 440\,D]$ with particle numbers between $N=3042$ and $24$,$336$.
For each density, we generated $5000$ independent configurations using the sampling method described in Ref.\,\cite{Vink2005} and determined the probability, $n(\lambda)$, of the occurence of wrapping clusters as a function of the connectivity. 
By standard practice, we take for the critical connectivities $\lambda_\pp$ the inflection points of a hyperbolic tangent, obtained by curve fitting $n(\lambda)$.\,\cite{BinderBook}
The error bars indicate the standard errors of the least-squares fit of $\lambda_\pp$.
We did not attempt an explicit finite-size scaling analysis,\,\cite{Finner2018} as it is non-trivial due to the dependence of the cluster shape distribution on the density and the proximity to the percolation threshold (see discussion below). 
Instead, we varied the box size in the indicated ranges and found the variation in the inflection points to be small.

\begin{figure}
	\includegraphics[width=\linewidth]{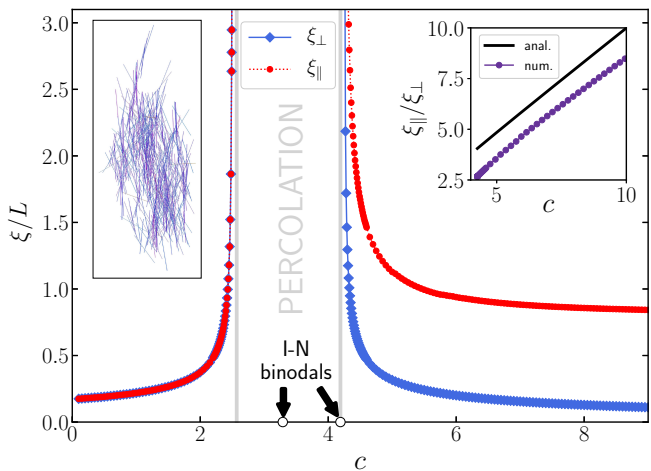}
	\caption{
		Correlation lengths parallel and perpendicular to the director (an arbitrary axis in the isotropic phase) obtained numerically, \textit{vs.} particle concentration for $\lambda=0.2\,D$ (regime C, see Figure \ref{fig:phasediagram}).
		Indicated are also the percolation thresholds in the isotropic and nematic phase.
		\textbf{Right inset:} analytical and numerical results for the cluster elongation in the nematic, for $\lambda=0.2\,D$.
		\textbf{Left inset:} snapshot of a sub-critical cluster from our MC simulations near percolation for $L/D=100$ at $c=3.72$ and $\lambda=0.16\,D$.
	}
	\label{fig:corrLengths}
\end{figure}

The advantage of our simulations is that we get direct information about the cluster shape, which in the isotropic phase is on average spherical but in the nematic is not, see the left inset to Figure \ref{fig:corrLengths}.
Clusters are much longer than they are wide, and oriented along the nematic director.
Our numerical prediction for the correlation lengths, indicated in the same Figure, confirms this.
Plotted are correlation lengths parallel and perpendicular to the director as a function of the dimensionless concentration $c$ for the connectivity $\lambda=0.2D\,$, for which we have percolation both in the isotropic and the nematic phase (regime C).
Also indicated are the phase gap and the percolation thresholds in both phases.
Again, in regime C, percolation in the isotropic phase occurs for concentrations \textit{above} the percolation threshold, while in the nematic it occurs \textit{below} the percolation threshold.
In the nematic, the correlation length along the director, $\xi_\parallel$, is much larger than that perpendicular to it, $\xi_\perp$, but both diverge at the same concentration.
The critical exponents are $1/2$, like in the isotropic phase, owing to the mean-field character of the problem in the limit of infinite aspect ratios.\,\cite{Bug1985}
In isotropic suspensions, where the director is an arbitrary axis, both lengths are equal, producing spherical clusters with
$\xi_\parallel/L=\xi_\perp/L=\sqrt{4c\lambda S/\pi D}/12$, 
exact in the slender rod limit.

According to our variational theory with the Gaussian approximation, we find in the nematic
$\xi_\parallel/L\sim\sqrt{\lambda S/3D}$
with $S$ the cluster size and
$\xi_\perp/L\sim\sqrt{5\pi\lambda S/48Dc^2}$.
This implies that, for large $c$, the parallel correlation length becomes a constant, while the one perpendicular to the director decreases as $1/c$.
The ratio $\xi_\parallel/\xi_\perp\sim4c/\sqrt{5\pi}$ measures the anisotropy of sub-critical clusters, which increases linearly with concentration.
Our numerical results support these conclusions, as the right inset to Figure \ref{fig:corrLengths} shows.
Strongly anisotropic cluster dimensions suggest that near the percolation threshold in a \textit{finite volume,} e.g., in a thin film or fiber, percolation might not persist in all directions.
This could aid the design of materials with strongly anisotropic conductivities or dielectric properties.\,\cite{Zamora2011}

A natural question is how sensitive our results are to ``imperfections'' such as deviations from the cylindrical particle shape and from monodispersity, which in practice are difficult to avoid.
What is known, is that the precise particle shape has very little effect on the percolation threshold in the isotropic phase.\,\cite{Drwenski2017} 
In contrast, polydispersity in length or width has been predicted to significantly affect the percolation threshold of both ideal and hard rods.\,\cite{Otten2009,Nigro2013,Finner2018}

Tortuosity in the form of a weakly helical shape can straightforwardly be incorporated into our theory for percolation in the (chiral) nematic.
Assume a helix of contour length $L$, width $D$, helical radius from the centerline $\delta$ and helical wave number $Q=2\pi/p$ with $p$ the molecular pitch.
(The sign of $Q$ and $p$ determines the handedness.)
This simply renormalizes the two-particle excluded volume by a factor $\nu=1-15Q^2\delta^2/16$, at least in the limit $|Q\delta|\ll1$ and $\delta\ll D$, and shifts the isotropic-nematic transition to larger filler fractions by a factor $\nu^{-1}$.\,\cite{Wensink2015}
Not surprisingly, the same renormalization applies to the contact volume, and the percolation threshold in the isotropic phase shifts upwards in the same fashion.
As the correction rescales both the excluded \textit{and} the contact volume, we find within our Gaussian variational theory that in the (chiral) nematic a weak helicity has no effect whatsoever on the percolation threshold, due to exact cancellation of the correction terms.
In fact, the effect of a weak tortuosity on the percolation diagram in Figure \ref{fig:phasediagram} can be scaled out by redefining the concentration as $\nu c$.

Whether polydispersity impacts upon the percolation threshold in the nematic as much as in the isotropic phase remains to be seen.
However, noting the deep relation between the osmotic compressibility and the cluster size in the context of connectedness percolation theory,\,\cite{Coniglio1977} we surmise that this is not the case.
The reason is that the osmotic pressure of a hard-rod nematic is an invariant of any length polydispersity, at least within the Gaussian approximation.\,\cite{Odijk1986review}
In fact, because the osmotic pressure of the nematic is proportional to the particle density, the compressibility is density-invariant.\,\cite{Odijk1986review}
This supports our finding that also the cluster size and the percolation threshold in the Gaussian approximation are constant.
Close to the nematic melting transition this approximation becomes less accurate, ultimately leading to a weak concentration dependence of the critical connectivity, as demonstrated in this Letter.


In summary, we have shown by means of connectedness percolation theory and Monte Carlo simulations that geometric percolation in the nematic liquid crystal of hard rods is fundamentally different from that in the isotropic phase.
In the nematic, the cluster size depends only very weakly on the particle loading and is, in essence, governed by the connectivity range, which in nanocomposites relates to the tunneling distance of electrons through the polymeric host, and in aqueous dispersions to the Debye length, if particles are charge stabilized.\,\cite{Vigolo2005}
This gives rise to four connectivity regimes defined by whether or not percolation is possible in either or both phases.
We also find that clusters in the nematic are strongly elongated along the director,
which may be used to produce materials with anisotropic properties.

S.\,P.\,F.\ and P.\,v.\,d.\,S.\ are funded by the European Union's Horizon 2020 research and innovation programme under the Marie Sk\l{}odowska-Curie grant agreement No 641839.

\bibliography{lit}
	
\end{document}